\documentclass[prl,amsmath,showpacs,twocolumn,letterpaper,showkeys]{revtex4}
\usepackage{times}
\usepackage{pstricks}
\usepackage{graphicx}

\usepackage{amsmath}
\usepackage{amssymb}
\usepackage{marvosym}
\usepackage{dsfont}
\usepackage{bezier,color,epsfig}

\begin{document}
\title{Collective Feshbach scattering of a superfluid droplet from a
  mesoscopic two-component Bose-Einstein condensate} 
\author{M. Grupp}
\author{G. Nandi}
\author{R. Walser}
\author{W. P. Schleich}
\affiliation{Universit\"at Ulm, Abteilung Quantenphysik, 
  D-89069 Ulm, Germany}
\date{\today} 

\begin{abstract}
  We examine the collective scattering of a superfluid droplet impinging on a
  mesoscopic Bose-Einstein condensate (BEC) as a target. The BEC consists of
 an atomic gas with two internal electronic states, each of which is trapped
  by a finite-depth external potential. An off-resonant optical laser field
  provides a localized coupling between the BEC components in the trapping
  region.  This mesoscopic scenario matches the microscopic setup for Feshbach
  scattering of two particles, when a bound state of one sub-manifold is
  embedded in the scattering continuum of the other sub-manifold. Within the
  mean-field picture, we obtain resonant scattering phase shifts from a linear
  response theory in agreement with an exact numerical solution of the real
  time scattering process and simple analytical approximations thereof. We
  find an energy-dependent transmission coefficient that is controllable via
  the optical field between 0 and 100\%.
\end{abstract}

\pacs{03.65.N, 03.75.Fi, 34.50} \keywords{Bose-Einstein condensation, BEC, 
scattering theory, Feshbach resonances} \maketitle


%
The natural way to investigate quantum objects is scattering. By selecting a
convenient physical ``stencil'' with a large interaction cross section, one
can probe the structure as well as the excitation properties of a target.  It
mostly happens that quantum objects are either microscopically small, like
atoms or nuclei, or they are embedded inside a solid state system like
electrons, electronic holes or Cooper pairs. Thus, they have an elusive
character, which usually shuns direct observation.
  
The generic response of a compound quantum object to bombardment with
projectiles is either individual, i.\thinspace{}e., by instantaneously
ejecting another single particle of the compound, or it is collective, when
after some transient period the target responds as a whole. Probably, the most
drastic instance of this collective behavior is nuclear fission, when a heavy
nucleus breaks up into large fragments -- unleashing large amounts of kinetic
energy.  Modeling the dynamics of the atomic nucleus as a classical liquid
drop \cite{hill53,greiner96a} gave a very intuitive interpretation of the
observed phenomenon.

Superconductivity in metals is another prominent collective effect.  Within
the Ginzburg-Landau theory \cite{abrikosov88}, one associates a collective
wave function with the Bose-condensed electronic Cooper-pairs, and a
hydrodynamic description is again successful. The quantum mechanical nature of
the fluid-like order parameter is usually discussed with the Josephson effect,
but Andreev-Saint-James reflection
\cite{deutscher05} 
is an equally interesting phenomenon and much more in line with collective
scattering theory that will be presented in the following.  This effect
explains the unusual electrical transport properties through a normal
metal-superconductor (N-S) junction. The occurrence of collective quantum
mechanical resonances can be explained by a conversion of normal conductor
electrons into hole-like excitations at the interface \cite{griffin71}.

The discovery of superfluidity in bosonic \cite{southwell,pethick02} and, most
recently, fermionic atomic gases  \cite{regal04,chin04,zwierlein04} was an
amazing achievement and is another manifestation of collective many-particle
physics.
Probably, the use of interatomic Feshbach scattering resonances has been the
most fruitful novel concept of the past few years. Today, they are the
universal tool to manipulate the binary interaction in an atomic gas
\cite{tiesinga92,Inouye98,cornish400} in real time and they paved the way to
fermionic superfluidity \cite{regal04,chin04,zwierlein04,kokkelmans501}.

\begin{figure}[b]
  \includegraphics[angle=-90,width=\columnwidth]{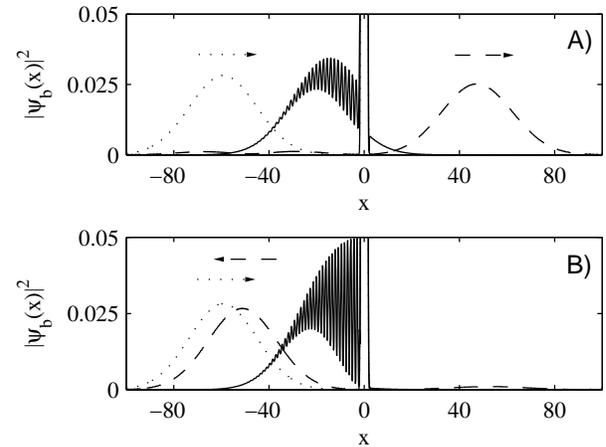}
  \caption{Collective Feshbach resonance: 
    complete transmission (A) or total refection (B) of a small coherent
    atomic wave-packet $\psi_b(x,t)$ with incident momentum $k_0=2.05$ (A) or
    $k_0=2.35$ (B), when scattering-off a stationary, two-component BEC,
    trapped around $-1<x<1$, with a maximal density $n_b(x)\approx 500$, which
    is off-scale. The dimensionless density $n_b(x,t)$ is depicted versus
    position $x$ (in natural units of the trap \cite{scale}) for three
    instants $t$: initially ($t_i$, dotted), on impact ($t_0$, solid) and
    finally ($t_f$, dashed). }
  \label{dynamic}
\end{figure}

%
In the present Letter, we will demonstrate that the microscopic physics of
binary Feshbach resonances can be also implemented at the mesoscopic level of
an atomic BEC, giving rise to collective Feshbach resonances as shown in
Fig.~\ref{dynamic}.  In particular, we will study the scattering properties of
a weak coherent perturbation on a two-component BEC confined in quasi
one-dimensional square-well potentials of finite-depth.  Based on a
Gross-Pitaevskii (GP) mean-field picture, we derive a two-component linear
response Bogoliubov theory \cite{tommasini03} for the scattering phases of the
continuum perturbations.  This is compared with the numerical simulation of
the nonlinear wave-packet propagation in real time. Finally, we can
explain the appearance of collective Feshbach resonances qualitatively from a
simple Thomas-Fermi approximation of the Bogoliubov excitations.
\begin{figure}[b]
  \includegraphics[angle=0,width=\columnwidth]{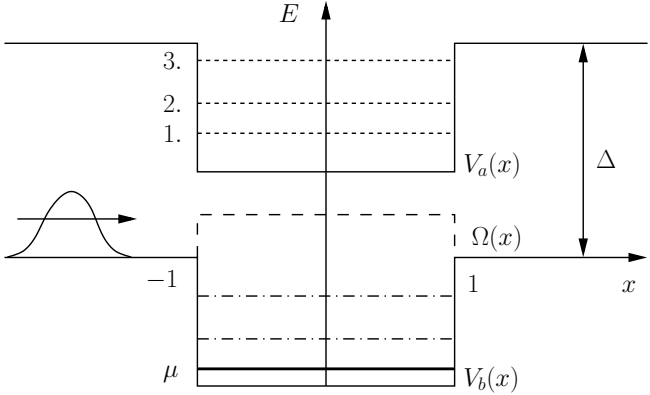}
  \caption{Schematic set-up of a trapped two-component BEC 
    with square-well potentials $V_a(x)$, $V_b(x)$ (solid) and a coupling
    laser beam $\Omega(x)$ (long dashed) versus position $x$ \cite{scale}.
    Scattering occurs in the two open (left/right) b-channels, while the
    a-channels are energetically closed with a threshold energy $\Delta$.  The
    chemical potential is $\mu$ (heavy solid line), the
    dashed-dotted lines show the bound excitations and the numbered
    quasi-bound energy levels (dashed) are responsible for the collective
    Feshbach resonances.}
  \label{squarewell}
\end{figure}

The dynamical response of a BEC has been the subject of intensive
investigations, during the last decade.  So far, laser light has been mostly
the method of choice to impart momentum onto a BEC, to excite collective modes
and to measure the dynamic structure factor \cite{ozeri05}.  However, light
has also been used indirectly to prepare colliding matter-wave packets that
have exhibited stimulated amplification as a result of their non-classical
bosonic nature \cite{vogels03}.  While the collision energy of these wave
packets was rather low in this setup (few times the speed of sound), more
recently, high energy, but ultra-cold thermal \cite{julienne04} and condensed
atomic clouds \cite{walraven04} of equal size have been used to measure the
single-particle collision cross section.  Theoretically, scattering of
identical particles off a scalar BEC was already considered in the limits of
low and high energies \cite{kuklov99,poulsen03}.  Scattering of vortices was
investigated in \cite{bijlsma00} (already relating it to Andreev-Saint-James
reflections \cite{deutscher05}) and a generalization of Levinson's theorem was
presented in \cite{brand03}.
 
Binary Feshbach scattering resonances cannot be described by a single channel
potential scattering but they require at least the interaction of two
different state manifolds. This was discussed first in nuclear scattering
theory by H.~Feshbach \cite{Feshbach1962b,tobocman61,friedrich} and in the
context of optical spectra of two-electron atoms by U.~Fano
\cite{Fano1961,Agarwal1984}.  Thus, we will assume that the quasi
one-dimensional condensate consists of atoms with two distinct internal
electronic states.  In the condensed phase, the system is described with order
parameters $\psi_a$ and $\psi_b$ that are coupled by an optical laser beam
$\Omega(x)$ with a large detuning $\Delta$.  Today, this can be achieved
experimentally with very prolate traps, which freeze out the transverse
degrees of motion, effectively.  The corresponding two-component GP equation
reads is also known from the interal Josephson effect \cite{williams199}
\begin{gather}
\label{GP}
\left[i \,\partial_t+\frac{1}{2}\,\partial^2_x-
\begin{pmatrix}
  V_a^{\text{GP}}(x) & \Omega(x) \\
  \Omega^\ast(x) &V_b^{\text{GP}}(x)
\end{pmatrix}\right] \,
\begin{pmatrix}
  \psi_a \\
  \psi_b
\end{pmatrix}
=0,\\
V_a^{\text{GP}}(x)=V_a(x) + g_{aa}\,|\psi_a(x)|^2+
g_{ab}\,|\psi_b(x)|^2,\nonumber\\
V_b^{\text{GP}}(x)= V_b(x) + g_{ab}\,|\psi_a(x)|^2+
g_{bb}\,|\psi_b(x)|^2,\nonumber
\end{gather}
where $V_a$ and $V_b$ are external trapping potentials. For definiteness, we
pick the square-well potentials as they lead to simple, analytically solvable
approximations. This setup is depicted in Fig.~\ref{squarewell}.  The coupling
constants $g_{aa}$, $g_{ab}=g_{ba}$, and $g_{bb}$ are proportional to the
self- and cross-component s-wave scattering lengths. For typical experimental
values of $^{87}$Rb see \cite{hall98}, but from the theoretical point of view
the choice of parameters is uncritical \cite{scale}.  In order to describe the
scattering of a superfluid droplet from the equilibrium BEC, we will determine
the linear response modes of the two-component system \cite{tommasini03}. The
stationary Bogoliubov ansatz with particle-like
$\boldsymbol{u}=(u_a,u_b)^\top$ and hole-like $\mathbf{v}=(v_a,v_b)^\top$
excitations is
\begin{equation*}
  \boldsymbol{\psi}(x,t)
  =
  e^{-i\,\mu\,t}
  [\boldsymbol{\psi}^{(0)}(x) + e^{-i\,\epsilon\,t} \mathbf{u}(\epsilon,x)
  +e^{i\,\epsilon\,t} \mathbf{v}^\ast(\epsilon,x)],
\end{equation*}
where $\mu$ is the chemical potential of the spinorial ground state
$\boldsymbol{\psi}^{(0)}=(\psi^{(0)}_a,\psi^{(0)}_b)^\top$ and $\epsilon$ is
the excitation energy.  This perturbation to Eq.~(\ref{GP}) yields the
four-dimensional Bogoliubov equations
\begin{gather}
  \label{bog2k}
  \left[\epsilon+ \frac{\sigma_3}{2}\,\partial^2_x
    -\begin{pmatrix}
      V^{\text{B}} & M \\
      -M^\ast& -V^{\text{B}\ast}
    \end{pmatrix}
  \right] \begin{pmatrix}
    \mathbf{u} \\ 
    \mathbf{v}
  \end{pmatrix}=0,\\
  M=
  \begin{pmatrix}
    g_{aa}\,{\psi_a^{(0)}}^{2} & g_{ab}\,\psi_a^{(0)}\psi_b^{(0)}\\
    g_{ab}\,\psi_a^{(0)}\psi_b^{(0)} & g_{bb}\,{\psi_b^{(0)}}^{2}
  \end{pmatrix},
  \quad
  \sigma_3=\begin{pmatrix}
    \mathds{1}& 0\\
    0 & -\mathds{1}
  \end{pmatrix},\nonumber\\ 
  V^{\text{B}}=
  \begin{pmatrix}
    V_a^{\text{HF}}(x)& \Omega(x)+g_{ab}\,\psi_a^{(0)}\psi_b^{(0)\ast} \\
    \Omega^\ast(x)+g_{ab}\,\psi_a^{(0)\ast}\psi_b^{(0)} & V_b^{\text{HF}}(x)
  \end{pmatrix},\nonumber\\
  V_a^{\text{HF}}(x)= V_a(x)+2\,g_{aa}\,|\psi_a^{(0)}|^2+
g_{ab}\,|\psi_b^{(0)}|^2-\mu,\nonumber\\
  V_b^{\text{HF}}(x)= V_b(x)+g_{ab}\,|\psi_a^{(0)}|^2+
2\,g_{bb}\,|\psi_b^{(0)}|^2-\mu.\nonumber
\end{gather}

\begin{figure}[h]
  \includegraphics[angle=-90,width=\columnwidth]{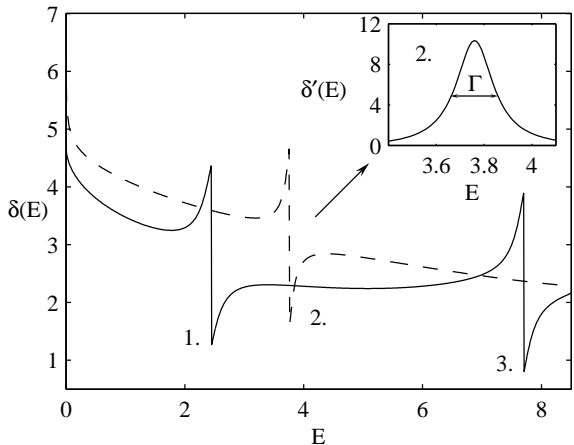}
  \caption{Collective Feshbach resonances in the scattering
    phases $\delta_e$ (solid) and $\delta_o$ (dashed) of a weak perturbation
    of a two-component BEC versus energy $E$. Note that the $\pi$-jumps of the
    collective Feshbach resonances occur at excitation energy
    $E_i=\mu+\epsilon_i$ of the even/odd quasi-bound Bogoliubov modes of
    Fig.~\ref{squarewell}. The inset shows the Lorentzian behavior of the
    phase derivative $\delta_o^\prime(E)$ close to the resonance as in
    Eq.~(\ref{bw}). For parameters see \cite{scale}.}
  \label{fig:phabog}
\end{figure}

The linear response energy matrix of Eq.~(\ref{bog2k}) has the same structure
as the well-known Bogoliubov equations in the case of a one-component
condensate. It is symplectic and has a real-valued spectrum of pairwise
positive and negative eigenvalues \cite{blaizot}.  Due to the finite-depth of
the trapping potentials, the spectrum supports only a finite number of bound
states and has a scattering continuum above a certain excitation energy.  This
is now the analogous situation as required for the two-particle Feshbach
scattering resonances. If quasi-bound Bogoliubov modes coincide in energy with
continuum modes, we will obtain a resonance behavior. These bound, positive
energy modes are depicted schematically in Fig.~\ref{squarewell}. For positive
energy solutions $\epsilon>0$, these resonances appear in the domain
$0<\epsilon+\mu=k^2/2=E < \Delta$. In this regime, the mode components
$u_a(\epsilon,x)$, $v_a(\epsilon,x)$ and $v_b(\epsilon,x)$ are localized on
the condensate and vanish exponentially for $|x|\rightarrow \infty$.  Only the
particle-like component $u_b(\epsilon,x)$ can propagate outside
\begin{equation}
\label{ub}
\lim_{x\rightarrow\pm\infty} u_b(\epsilon,x)=\cos(kx \pm \delta_e),  \text{ or } \sin(kx \pm \delta_o).
\end{equation}
Due to the reflection symmetry of the trapping potentials, the excitation can
also be characterized by parity and we define even and odd phase shifts
$\delta_e$ and $\delta_o$, respectively.  These scattering phases have been
evaluated numerically from Eqs.~(\ref{GP},\ref{bog2k},\ref{ub}) and are
displayed in Fig.~\ref{fig:phabog}. The analogy to the phenomenon of Feshbach
resonances stands out very clearly.  According to the Breit-Wigner
parameterization of an isolated resonance \cite{friedrich}, one finds
approximately
\begin{equation}
  \label{bw}
  \delta^{\text{res}}(E)=\arctan\left(\frac{\Gamma/2}{E_R-E}\right),\quad
  \frac{2}{\Gamma}=\left.\frac{d\delta^{\text{res}}}{dE}\right|_{E=E_R}.
\end{equation}
In turn, one can determine the resonance energy $E_R$ and width $\Gamma$
analytically from the poles of the $S$-matrix in the vicinity of a Feshbach
resonance \cite{Newton80,grupp05,forthcoming}.

\begin{figure}[h]
  \includegraphics[angle=-90,width=\columnwidth]{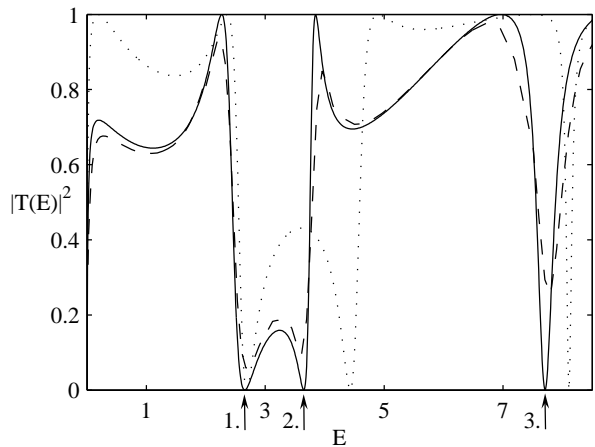}
  \caption{Transmission coefficient of a weak coherent perturbation 
    in the b-component of the trapped BEC versus energy $E$ from the linear
    response Bogoliubov calculation (solid).  The dashed line is the
    transmission coefficient obtained from propagating a Gaussian wave-packet
    in the nonlinear GP Eq.~(\ref{GP}), in real time. The dotted line
    represents a crude Thomas-Fermi approximation.  Feshbach resonances are
    seen clearly in all three curves superimposed on a background caused by
    potential scattering.}
  \label{fig:spektrum}
\end{figure}

From the phases shifts, one can obtain all scattering information, like
reflection $R(E)$ or transmission amplitudes $T(E)$, by considering a causal
wave $u_b^{(+)}(\epsilon,x)$, propagating to the right ($k>0$),
\begin{equation}
  \begin{split}
    \lim_{x\rightarrow-\infty}&u_b^{(+)}(\epsilon,x)=e^{ikx}+R\,e^{-ikx}\\
    &=e^{i\delta_e}\cos(kx-\delta_e)+ie^{i\delta_o}\sin(kx-\delta_o),\\
    \lim_{x\rightarrow\infty}&u_b^{(+)}(\epsilon,x)=T\,e^{ikx}\\
    &=e^{i\delta_e}\cos(kx+\delta_e)+ie^{i\delta_o}\sin(kx+\delta_o).
  \end{split}
\end{equation}
The real-valuedness of the scattering phases of Fig.~\ref{fig:phabog} implies a
current conservation $|R(E)|^2+|T(E)|^2=1$, and we can write the transmission
coefficient in terms of the phases
\begin{equation}
  \label{transmission}
  \left|T(E)\right|^2=\cos^2[\delta_e(E)-\delta_o(E)].
\end{equation}
This transmission coefficient is shown in Fig.~\ref{fig:spektrum}. In the
vicinity of the resonance energies $E\approx \mu+\epsilon_i$, it changes
rapidly between 0 and 100\%.

In order to isolate the essential physical mechanism responsible for the
collective resonance behavior, we approximate the GP Eq.~(\ref{GP}) in the
Thomas-Fermi limit.  Then $\psi_{\text{TF}}^{(0)}(x)$ is constant within the
square-wells and vanishes exactly elsewhere. In an additional approximation,
we disregard the matrix $M$ in the Bogoliubov self-energy Eq.~(\ref{bog2k}).
In this limit, the particle and hole excitations are decoupled and
$\mathbf{u}$ satisfies a two-component Schr\"odinger equation
\begin{gather}
  \label{sglike}
  (\epsilon+ \frac{\partial^2_x}{2} - V^{\text{B}})\mathbf{u}=0.
\end{gather}
Due to the TF approximation, this equation has again a square-well character
and the solution can be found analytically \cite{tobocman61,forthcoming}.
Sparing the details of the calculation of the scattering phases, we simply
present the results in the dotted line in Fig.~\ref{fig:spektrum}. A good
qualitative agreement needs to be acknowledged, while there are obviously
shifts in the resonance energies that are not accounted for in this simple
approximation scheme.

In a final step of the analysis of the collective Feshbach resonance, we have
also performed a numerical simulation of the nonlinear, time-dependent GP
Eq.~(\ref{GP}).  We propagated an incident traveling Gaussian wave packet on
top of the stationary BEC solution depicted in Fig.~\ref{dynamic},
\begin{equation}
  \psi_b(x,t=0)=\psi_b^{(0)}(x)+\sqrt[4]{\frac{\delta N^2}{\pi
      \sigma^2}}\,e^{-\frac{(x-x_0)^2}{2\sigma^2}+i k_0 \,x}.
\end{equation}
It can be seen in Fig.~\ref{dynamic} that the b-component of the ground state
solution $\psi_b^{(0)}$ is well localized in the trap center and has initially
no overlap with the weak Gaussian perturbation if $\delta N\ll N$ and $x_0 \ll
-1$. The initial momentum $k_0>0$ of the wave was varied to cover the energy
range in Fig.~\ref{fig:spektrum}. In order to resolve the resonance structure,
one needs a small momentum spread ($\sigma=20$) and we find that the
wave-packet transmission spectrum matches the linear response approach very
well. We show two instances of the propagation of an initial wave packet. In
Fig.~\ref{dynamic}A), the momentum $k_0\approx 2.05$ corresponds to an energy
below the resonance energy, marked as $E_1$ in Fig.~\ref{fig:spektrum}, which
leads to full transmission.  In contrast, Fig.~\ref{dynamic}B) shows the total
reflection of the wave-packet if the incident momentum $k_0\approx 2.35$,
corresponds exactly to the resonance energy $E_1$.

In conclusion, we have identified collective Feshbach resonances in a trapped
two component BEC. These resonances do not require a binary Feshbach
resonance to modify the binary interaction, but quasi-bound Bogoliubov modes
in the linear response spectrum. We have shown that this can be achieved
easily with help of an optical laser beam. In contrast to the microscopic
binary Feshbach resonance, this quantum mechanical phenomenon exists on a
mesoscopic, possibly macroscopic scale and can be used to control the
transmission of matter-waves between 0 and 100\%. For conceptual simplicity, we
have investigated a quasi one-dimensional geometry with square-well
potentials. None of this is important for the effect and it can
be implemented in various experimental configurations. However, studying this
resonance in the mean-field picture can be only a first step and a proper
inclusion of the thermal cloud is still lacking \cite{walser999,walser04}.

We thank E. Kajari for fruitful discussions and gratefully acknowledge
financial support by the SFB/TR 21 {\em ``Control of quantum correlations in
tailored matter''} funded by the Deutsche Forschungsgemeinschaft (DFG).

\bibliographystyle{apsrev}
\bibliography{bec,MyPublications,grupp05}

\begin{thebibliography}{37}
\bibitem[{sca()}]{scale}
\bibinfo{note}{The physical length, energy and time scales are the half-size of
  the square-well box $a$, the energy $E=\hbar \omega =\hbar^2/m a^2$ and a
  time scale $t=2 \pi/\omega$. We have solved the model for the collective
  Feshbach resonances for $N=1000$ particles, coupling constants
  $g_{aa}=g_{ab}=g_{bb}=0.008$, potential depths $V_a=V_b=-15$, a Rabi
  frequency $\Omega=7$, a detuning $\Delta=8.5$ and get for the chemical
  potential $\mu\approx-13.5$.}
\end{thebibliography}
\end{document}